\documentclass[aps,a4paper,twocolumn,preprintnumbers,showpacs,showkeys,
nofootinbib,superscriptaddress]{revtex4}
\usepackage{amssymb,amsmath,amsfonts,amsthm,graphicx}

\graphicspath{{./Fig/}}                 

\newcommand{\beq}{\begin{equation}}
\newcommand{\eeq}{\end{equation}}
\newcommand{\bea}{\begin{eqnarray}}
\newcommand{\eea}{\end{eqnarray}}
\newcommand{\beas}{\begin{eqnarray*}}
\newcommand{\eeas}{\end{eqnarray*}}
\newcommand{\ii}{\ensuremath{\text{i}}}

\newcommand{\psibar}{\ensuremath{\overline\psi}}
\newcommand{\chibar}{\ensuremath{\overline\chi}}
\newcommand{\pbp}{\ensuremath{\psibar\psi}}

\newcommand{\vev}[1]{\ensuremath{\left<#1\right>}}
\newcommand{\Fig}[1]{Fig.\,\ref{#1}}
\newcommand{\Tab}[1]{Table\,\ref{#1}}
\newcommand{\Sec}[1]{Section\,~\ref{#1}}
\newcommand{\Eq}[1]{Eq.\,(\ref{#1})}
\newcommand{\re}{\operatorname{\mathfrak{Re}}}
\newcommand{\tr}{\operatorname{Tr}}

\begin{document}

\title{Landau gauge gluon and ghost propagators from lattice QCD
with $N_f=2$ twisted mass fermions at finite temperature}
\author{R.~Aouane}
\affiliation{Humboldt-Universit\"at zu Berlin, Institut f\"ur Physik, 
12489 Berlin, Germany}
\author{F.~Burger}
\affiliation{Humboldt-Universit\"at zu Berlin, Institut f\"ur Physik, 
12489 Berlin, Germany}
\author{E.-M.~Ilgenfritz}
\affiliation{Humboldt-Universit\"at zu Berlin, Institut f\"ur Physik, 
12489 Berlin, Germany}
\affiliation{Joint Institute for Nuclear Research, VBLHEP, 141980
  Dubna, Russia} 
\author{M.~M\"uller-Preussker}
\affiliation{Humboldt-Universit\"at zu Berlin, Institut f\"ur Physik, 
12489 Berlin, Germany}
\author{A.~Sternbeck}
\affiliation{Universit\"at Regensburg, Institut f\"ur Theoretische Physik, 
93040 Regensburg, Germany}

\preprint{HU-EP-12/45}

\date{March 13, 2013}

\begin{abstract}
We investigate the temperature dependence of the Landau gauge gluon
and ghost propagators in lattice QCD with two flavors of maximally
twisted mass fermions. For these propagators we provide and analyze
data which corresponds to pion mass values between 300 and
500\,MeV. For the gluon propagator we find that both the 
longitudinal and transversal component change smoothly in the
crossover region, while the ghost propagator exhibits 
only a very weak temperature dependence. For momenta between 0.4 and 3.0\,GeV
we give a parametrization for our lattice data. It may serve as input
to studies which employ continuum functional methods.
\end{abstract}



\maketitle

\section{Introduction}
\label{sec:introduction}

Hadronic matter may experience a phase transition or crossover from a
confined phase, with broken chiral symmetry, to a deconfined
chirally symmetric phase. The latter phase is characterized by a state
called {\it quark gluon plasma}. It is widely believed this state of
matter has been passed in the early universe, and efforts are undertaken 
to reproduce it experimentally in heavy-ion collisions as in present
collider experiments at RHIC (BNL) or with the ALICE and CMS detectors 
at LHC (CERN).

The possible existence of such a phase transition was first discussed
\cite{Cabibbo:1975ig} in the context of Hagedorn's thermodynamic model
\cite{Hagedorn:1968zz,Hagedorn:1970gh} as a way to evade the
consequence of a maximal hadronic temperature. First numerical evidence
has been found already in the early days of lattice QCD (LQCD)
\cite{McLerran:1980pk,Kuti:1980gh}. In fact, the lattice formulation
offers an ab-initio approach to study such aspects of QCD
nonperturbatively. This remains true as long as the chemical potential
is small compared to the temperature. Over the last decade immense
computational resources were dedicated to reach at a consistent
picture of QCD at finite temperature and zero chemical potential (for
recent reviews see~\cite{Kanaya:2010qj,Levkova:2012jd,Philipsen:2012nu,
Lombardo:2012}).

LQCD, however, is not the only framework to tackle nonperturbative
problems of QCD at zero or non-zero temperature. Powerful continuum
functional methods exist as well, like for instance in the context of
Dyson-Schwinger (DS) equations
~\cite{vonSmekal:1997vx,Hauck:1998fz,Roberts:2000aa,Maris:2003vk} or
functional renormalization group (FRG) equations
~\cite{Gies:2002af,Pawlowski:2003hq,Braun:2009gm}, 
which also allow to address such problems.

Within these frameworks (see, e.g., the recent reviews
\cite{Fischer:2008uz,Boucaud:2011ug}) the Landau gluon and ghost
propagators appear---together with the corresponding vertices---as the
main building blocks in the formulation of the DS or FRG equations;
they constitute part of the solutions of the latter. These functional
methods though come with a potential source of error: To solve the
(infinite) tower of equations it has to be truncated appropriately,
and so the solutions beyond the far-infrared regime, depend on how the
truncations are done, in particular in the momentum range around
$O(1)$ GeV. Therefore, independent information, at best from first
principles, is welcome to improve these (unavoidable) truncations.

LQCD calculations allow to provide the Landau gluon and ghost
propagators in an ab-initio way. The available momentum range, however,
is restricted from above by the lattice spacing and from below by the
available lattice volume (up to a further uncertainty related to
so-called Gribov copies \cite{Bornyakov:2008yx,Bornyakov:2010nc}). 
Despite these restrictions, an impressive amount of data has been 
produced over the last years for these propagators at zero (see, e.g.,
\cite{Bornyakov:2009ug,Bogolubsky:2009dc} and references therein) and
non-zero temperature 
\cite{Heller:1995qc,Heller:1997nqa,Cucchieri:2000cy,Cucchieri:2001tw, 
Cucchieri:2007ta,Sousa:2007zzb,Maas:2009fg,Bornyakov:2009ug,Bornyakov:2010nc,
Fischer:2010fx,Bornyakov:2011jm,Cucchieri:2011di,Aouane:2011fv,Maas:2011ez,
Cucchieri:2012nx,Cucchieri:2012gb}. This data has allowed for a variety of
cross-checks with corresponding results from the continuum functional
methods. 

Conventional lattice calculations of QCD at finite temperature often
employ the Polyakov loop and the chiral condensate to probe for the
(de)confinement and chiral phase transition, respectively. In
recent years it turned out that these and other observables can be also
calculated from DS and FRG equations, where the Landau gauge
gluon and ghost propagators are used as input
\cite{Braun:2007bx,Fischer:2008yv,Fischer:2009wc,Fischer:2010fx, 
Fischer:2011pk,Fischer:2012vc,Fister:2011um,Fister:2011uw}. 
Even an extrapolation to the notoriously difficult regime of  
nonzero chemical potential seems to be possible \cite{Fischer:2011mz,
Lucker:2011dq}. Appropriate lattice data for the propagators close
to the continuum limit is therefore essential to assist those efforts. 

In a recent study \cite{Aouane:2011fv} we have provided such
data for quenched QCD (see, e.g., \cite{Fukushima:2012qa}
for a first application). We could show that the longitudinal 
(electric) component of the gluon propagator may be used to probe the
thermal phase transition of pure $SU(3)$ gauge theory. Similar was
shown in \cite{Maas:2011ez}. The present article further complements
the available lattice data with finite-temperature data for the Landau
gauge gluon and ghost propagators from LQCD with $N_f=2$ dynamical  
fermion flavors. To the best of our knowledge, there are only two studies which
have provided such data in the past \cite{Furui:2006py,Bornyakov:2011jm}.

For our study we adopt a lattice formulation that employs the
Symanzik-improved gauge action for the gluonic field and the twisted
mass Wilson-fermion action for the fermionic part. The latter ensures
an automatic $O(a)$ improvement provided the Wilson $\kappa$-parameter
is tuned to maximal twist (for further details we refer to
\cite{Shindler:2007vp,Urbach:2007rt}). Our data is based on the vast
set of gauge field configurations that has been generated by the tmfT
Collaboration. The tmfT Collaboration has explored the complicated
phase structure of the theory \cite{Ilgenfritz:2009ns} and is still
investigating the smooth crossover region from the confining and
chirally broken regime at low temperature $T$ to the deconfinement and
chirally restored phase at high $T$ \cite{Burger:2010ag,Burger:2011zc}. 
These studies are restricted so far to pseudo-scalar meson (pion)
masses from $300$ MeV to $500$ MeV.
To set the scale we use results at $T=0$ of the European Twisted Mass 
(ETM) Collaboration \cite{Baron:2009wt}. Specifically, our data for the 
gluon and ghost propagators covers the whole crossover regime at three 
pion mass values: $m_\pi \simeq 316$, 398 and 469\,MeV. At these values 
the crossover regime is characterized by a very smooth behavior of the 
chiral condensate and the Polyakov loop as well as their susceptibilities. 
Moreover, one observes for these settings the breakdown of chiral symmetry
and the deconfinement phase transition occur at slightly 
different temperatures $T_c = T_{\chi}$ and $T_{\mathrm{deconf}}$, 
respectively, in agreement with observations reported in 
\cite{Borsanyi:2010bp}.%
\footnote{Detailed results are presented in a recent update 
to \cite{Burger:2011zc}. We thank the tmfT Collaboration for 
providing us their $T_c$-data prior to publication.
}

The paper is organized as follows. In \Sec{sec:parameters} we 
give all lattice parameters and outline the setup of our Monte Carlo 
simulations. \Sec{sec:def_propagators} recalls the definitions of the 
gluon and ghost propagators on the lattice in the Landau gauge. 
Data and fits for various temperature and pion mass values are presented in
\Sec{sec:results_propagators}. Conclusions are drawn in
\Sec{sec:conclusion}.   

\section{Lattice action and simulation parameters}
\label{sec:parameters}

Our study is based on gauge field configurations provided by the 
tmfT Collaboration. These configurations were generated on a 
four-dimensional periodic lattice of spatial linear size $N_\sigma=32$ 
and a temporal extent of $N_\tau=12$ for a mass-degenerate doublet of twisted
mass fermions, cf.\ the review in Ref.~\cite{Shindler:2007vp}. The corresponding
gauge action is the tree-level Symanzik improved action defined as 
\bea \label{sym_action}
S_{G} = \beta \sum_{x}&&\left[c_{0} \sum_{\mu < \nu}
        (1-\frac{1}{3}\,\re\,\tr\,U^{1\times1}_{x\mu\nu}) \right.  \nonumber \\
                      &&\left.+ c_{1} \sum_{\mu \neq \nu}
        (1-\frac{1}{3}\,\re\,\tr\,U^{1\times2}_{x\mu\nu})\right] \,,\\
\eea
with $\beta= 2N_c / g_0^2$, $c_{1}=-\frac{1}{12}$, $c_{0}=1-8\,c_{1}$ and
$g_0$ being the bare 
coupling constant. $U^{1\times1}_{x\mu\nu}$ represents quadratic and
$U^{1\times2}_{x\mu\nu}$ rectangular Wilson loops built from the
link variables $U_{x\mu} \in SU(3)$. One important feature 
of this gauge action is its inherent $O(a)$ improvement. For more 
details see Refs.~\cite{Symanzik:1983dc, Symanzik:1983gh}. 
The Wilson fermion action with an additional parity-flavor symmetry 
violating improvement term reads \cite{Frezzotti:2000nk}
\bea
S_F[U,\psi,\psibar] = \hspace*{4.5cm} \nonumber \\ 
 \sum_x \chibar(x)\left( 1 -\kappa D_W[U] 
 + 2\ii\kappa a\mu_0\gamma_5\tau_3\right)\chi(x) \; ,
\label{twist_action}
\eea
where the Pauli matrix $\tau_3$ acts 
in flavor space and the fermionic fields are expressed in terms 
of the twisted basis $\{\chibar,\chi\}$, which is related to the 
physical fields $\{\psibar,\psi\}$ by
\beq
\psi = \frac{1}{\sqrt{2}}(1+\ii\gamma_5\tau_3)\chi \quad \text{and} \quad 
\psibar = \chibar \frac{1}{\sqrt{2}}(1+\ii\gamma_5\tau_3)\;.
\eeq
The Wilson covariant derivative acts on these as
\begin{eqnarray}
D_W[U]\psi(x)&=&\sum_{\mu} ((r-\gamma_{\mu})U_{\mu}(x)\psi(x+\hat{\mu}) 
\nonumber \\
&&(r+\gamma_{\mu})U_{\mu}^{\dagger}(x-\hat{\mu})\psi(x-\hat{\mu})) \;,
\end{eqnarray}
and the quark mass is set by the twisted mass parameter $\mu_0$ and the hopping
parameter $\kappa = (2am_0 + 8r)^{-1}$, parameterizing the untwisted bare
quark mass component. Here $a$ is the lattice spacing and $r=1$. 
Note that for any finite $\beta$ the value for $\kappa$ gets corrections
through mass renormalization. 

It must be noted that maximal twist is accomplished by tuning the hopping 
parameter to its critical value $\kappa_c$, where the untwisted theory 
would become massless. At maximal twist one achieves an automatic 
$O(a)$-improved fermion formulation \cite{Frezzotti:2003ni}. 
The hopping parameter entering the simulation is based on a set of 
$\beta$-values for which $\kappa_c(\beta)$ was provided by the
ETM Collaboration \cite{Baron:2009wt}. For intermediate $\beta$ the 
corresponding $\kappa_c$ values are obtained through an 
interpolation as described in Ref.~\cite{Burger:2011zc}. 
The bare twisted mass parameter $a \mu_0$ has been adjusted such as
to keep the physical pion mass constant along our scans in the bare inverse
coupling $\beta$.

As usual in finite temperature QCD, the imaginary time extent $N_{\tau}$
corresponds to the inverse temperature $T^{-1}=N_{\tau} a(\beta)$. To quote
it in physical units we use interpolated---as well as slightly
extrapolated---data for the lattice spacing reported for $\beta=3.90$, 4.05
and 4.20 by the ETM Collaboration~\cite{Baron:2009wt} (see also
Ref.~\cite{Burger:2011zc}). We restrict our analysis to lattice spacings
$a<0.09$ fm. 

For the reader's convenience all parameters, like $\beta$, the corresponding
lattice spacings, temperatures, pion masses, the number of independent
configurations and other relevant values are collected in \Tab{tab:numbers}.
Note that there, for definiteness, ``pion mass'' corresponds to the charged
pion. In \Tab{tab:tc} we provide also the respective pseudo-critical couplings 
$\beta_c$ and the corresponding temperatures $T_\chi$ and $T_\mathrm{deconf}$
for the three pion mass values we use. These temperatures were obtained
from fits around the maxima of the chiral susceptibility $\sigma^2_{\pbp}$ and
from the behavior of the (renormalized) Polyakov loop $\vev{\re(L)}_R$,
respectively 
(see the revised version of Ref.~\cite{Burger:2011zc}).

\begin{table*}
{\scriptsize
\mbox{
 \setlength{\tabcolsep}{3.0pt}
\begin{tabular}{|c|c|c|c|c|c|}
\hline
$m_{\pi}~\mathrm{[MeV]}$ &~$T~\mathrm{[MeV]}$ &~$a~\mathrm{[fm]}$ 
                                    &~$r_0/a$ &~$r_0\cdot T$ &~$n_{conf}$  \\
\hline
316(16) & 187 & 8.77(47)$\cdot10^{-2}$ & 4.81 & 0.40 & 293 \\ 
316(16) & 199 & 8.25(22)$\cdot10^{-2}$ & 5.17 & 0.43 & 299 \\ 
316(16) & 215 & 7.65(13)$\cdot10^{-2}$ & 5.63 & 0.47 & 255 \\
316(16) & 222 & 7.39(12)$\cdot10^{-2}$ & 5.84 & 0.49 & 273 \\
316(16) & 225 & 7.31(12)$\cdot10^{-2}$ & 5.91 & 0.49 & 151 \\
316(16) & 228 & 7.22(11)$\cdot10^{-2}$ & 5.98 & 0.50 & 250 \\
316(16) & 230 & 7.14(11)$\cdot10^{-2}$ & 6.05 & 0.50 & 113 \\
316(16) & 235 & 6.98(11)$\cdot10^{-2}$ & 6.19 & 0.52 & 290 \\
\hline\hline
398(16) & 193 & 8.51(32)$\cdot10^{-2}$ & 4.99 & 0.42 & 159 \\  
398(16) & 199 & 8.25(22)$\cdot10^{-2}$ & 5.17 & 0.43 & 173 \\
398(16) & 215 & 7.65(13)$\cdot10^{-2}$ & 5.63 & 0.47 & 209 \\ 
398(16) & 228 & 7.20(11)$\cdot10^{-2}$ & 6.00 & 0.50 & 198 \\                      
398(20) & 236 & 6.98(11)$\cdot10^{-2}$ & 6.19 & 0.52 & 156 \\
398(16) & 241 & 6.82(10)$\cdot10^{-2}$ & 6.34 & 0.53 & 150 \\
398(20) & 246 & 6.69(10)$\cdot10^{-2}$ & 6.46 & 0.54 & 271 \\
398(20) & 248 & 6.62(10)$\cdot10^{-2}$ & 6.53 & 0.54 & 226 \\
398(20) & 254 & 6.47(10)$\cdot10^{-2}$ & 6.68 & 0.56 & 113 \\
\hline\hline
469(24) & 222 & 7.42(12)$\cdot10^{-2}$ & 5.81 & 0.48 & 146 \\
469(24) & 228 & 7.20(11)$\cdot10^{-2}$ & 6.00 & 0.50 & 348 \\
469(24) & 235 & 6.98(11)$\cdot10^{-2}$ & 6.19 & 0.52 & 120 \\
469(24) & 243 & 6.77(10)$\cdot10^{-2}$ & 6.39 & 0.53 & 210 \\
469(24) & 247 & 6.67(10)$\cdot10^{-2}$ & 6.48 & 0.54 & 250 \\
469(24) & 250 & 6.57(10)$\cdot10^{-2}$ & 6.58 & 0.55 & 256 \\
469(24) & 254 & 6.47(10)$\cdot10^{-2}$ & 6.68 & 0.56 & 152 \\
469(24) & 258 & 6.38(10)$\cdot10^{-2}$ & 6.78 & 0.56 & 150 \\
469(24) & 266 & 6.19(12)$\cdot10^{-2}$ & 6.98 & 0.58 & 200 \\
\hline
\end{tabular}
}
\mbox{
\setlength{\tabcolsep}{1.0pt}
\begin{tabular}{|c|c|c|c|}
\hline
$\beta$& $\kappa_c(\beta)$ & $a\cdot\mu_0(\beta)$ \\
\hline
3.8400& 0.162731 & 0.00391 \\ 
3.8800& 0.161457 & 0.00360 \\ 
3.9300& 0.159998 & 0.00346 \\
3.9525& 0.159385 & 0.00335 \\
3.9600& 0.159187 & 0.00331 \\
3.9675& 0.158991 & 0.00328 \\
3.9750& 0.158798 & 0.00325 \\
3.9900& 0.158421 & 0.00319 \\
\hline\hline
3.8600& 0.162081 & 0.00617 \\ 
3.8800& 0.161457 & 0.00600 \\
3.9300& 0.159998 & 0.00561 \\
3.9700& 0.158927 & 0.00531 \\                                
3.9900& 0.158421 & 0.00517 \\
4.0050& 0.158053 & 0.00506 \\
4.0175& 0.157755 & 0.00498 \\
4.0250& 0.157579 & 0.00493 \\
4.0400& 0.157235 & 0.00483 \\
\hline\hline
3.9500& 0.159452 & 0.00779 \\
3.9700& 0.158926 & 0.00752 \\
3.9900& 0.158421 & 0.00738 \\
4.0100& 0.157933 & 0.00718 \\
4.0200& 0.157696 & 0.00708 \\
4.0300& 0.157463 & 0.00699 \\
4.0400& 0.157235 & 0.00689 \\
4.0500& 0.157010 & 0.00680 \\ 
4.0700& 0.156573 & 0.00662 \\
\hline
\end{tabular}
}
\mbox{
 \setlength{\tabcolsep}{4.0pt}
\begin{tabular}{|c|c|c|}
\hline
$\tilde{Z}_{T}$~&~$\tilde{Z}_{L}$~& $\tilde{Z}_J$ \\
\hline
0.6380(80) & 0.6264(108) & 0.66862(32) \\
0.6208(34) & 0.6139(50)  & 0.66939(10) \\
0.6117(70) & 0.6116(103) & 0.67264(14) \\
0.6156(59) & 0.6122(84)  & 0.67252(15) \\
0.6151(71) & 0.6101(102) &             \\
0.6120(63) & 0.6079(114) & 0.67388(14) \\
0.6146(77) & 0.5982(126) &             \\
0.6092(71) & 0.6118(97)  & 0.67536(14) \\
\hline\hline
0.6191(64) & 0.6147(98)  & 0.66730(21) \\ 
0.6202(54) & 0.6192(76)  &             \\
0.6076(59) & 0.6080(84)  & 0.67005(23) \\  
0.6087(59) & 0.6119(89)  &             \\
0.6075(139) & 0.6090(231)& 0.67293(13) \\
0.6156(87) & 0.6112(115) & 0.6784(60) \\
0.6036(78) & 0.6063(101) & 0.67504(16) \\
0.6001(64) & 0.6005(94)  & 0.67517(11) \\
0.6029(119) & 0.6167(178)& 0.67519(26) \\
\hline\hline
0.6121(55) & 0.6020(82)  & 0.67142(16) \\
0.6116(80) & 0.6024(110) & 0.67128(14) \\
0.6098(70) & 0.6041(102) & 0.67271(22) \\
0.6086(54) & 0.6093(73)  & 0.67322(12) \\
0.5947(54) & 0.5927(77)  & 0.67147(14) \\
0.6013(72) & 0.6017(101) & 0.67388(14) \\
0.6033(80) & 0.6028(121) & 0.67353(16) \\
0.5971(62) & 0.6072(91)  & 0.67485(17) \\
0.5972(143) & 0.6119(195)& 0.67829(32) \\
\hline
\end{tabular}
 }
\caption{The pion mass values, the temperature $T$, both in $\mathrm{MeV}$, 
the lattice spacing $a$ in $\mathrm{fm}$, the chirally extrapolated Sommer 
scale $r_0$ \cite{Sommer:1993ce}, $r_0~T$, and the number 
$n_{conf}$ of independent configurations used for the analysis are shown 
in the left subtable for all simulation ensembles. \\
In the central subtable the values of the inverse bare coupling $\beta$
used in the simulations, the critical hopping parameter $\kappa_c$ and the 
bare twisted mass $\mu_0$ are additionally shown. 
The spatial ($N_{\sigma}=32$) and temporal ($N_{\tau}=12$) sizes are the
same for all ensembles. The number $n_{copy}$ of gauge copies was fixed to 1. \\ 
In the right subtable the renormalization factors for the transverse and 
longitudinal gluon as well as for the ghost dressing function 
denoted by $\tilde{Z}_T$, $\tilde{Z}_L$ and $\tilde{Z}_J$, respectively
(see the text), are given for the renormalization scale $\mu=2.5$ GeV. 
} 
\label{tab:numbers}
}
\end{table*}

\begin{table}
\centering
     \begin{tabular}{r@{\quad}c@{\quad}c@{\quad}c}\hline\hline
      tmfT ensemble & A12        	& B12       & C12\\
      \hline
      $m_\pi\, [\text{MeV}]$      & $316(16)$  	& $398(20)$     & $469(24)$  \\ 
      $\beta_c$ from $\sigma^2_{\pbp}$   & $3.89(3)$  & $3.93(2)$ & $3.97(3)$ \\
      $T_\chi\, [\text{MeV}]$              & $202(7)$   & $217(5)$  & $229(5)$ 
\\
      $\beta_c$ from $\vev{\re(L)}_R$  & -- & $4.027(14)$ &  $4.050(15)$\\  
      $T_\mathrm{deconf}\, [\text{MeV}]$   & -- & $249(5)$  & $258(5)$ \\
  \hline\hline
    \end{tabular}
\caption{Extracted (pseudo-) critical couplings $\beta_c$ and the 
crossover-temperatures $T_\chi$ and $T_\mathrm{deconf}$ 
for the three ensembles A12, B12, and C12. Corresponding pion masses (from
\cite{Burger:2011zc}) are listed in the first row. Ensemble names indicate
$N_\tau=12$.}
\label{tab:tc}
\end{table}

\section{The gluon and ghost propagators on the lattice}
\label{sec:def_propagators}

Gluon and ghost propagators are gauge dependent quantities. As in
Ref.~\cite{Aouane:2011fv} we focus on  Landau gauge and therefore have to
transform the (unfixed) tmfT gauge ensemble until it satisfies the corresponding
gauge condition. In differential form it reads
\beq 
 \nabla_{\mu}A_{\mu}=
\sum_{\mu=1}^{4} \left(A_{\mu}(x+\hat{\mu}/2) - A_{\mu}(x-\hat{\mu}/2) \right) = 0
\label{eq:gaugecondition}
\eeq
with the lattice gauge potentials
\beq
 A_{\mu}(x+\hat{\mu}/2)=
               \frac{1}{2iag_{0}}(U_{x\mu}-U_{x\mu}^{\dagger})\mid_{traceless}\,.
\label{eq:potential}
\eeq
To render the link variables satisfying this condition one maximizes the gauge
functional
 \beq
F_{U}[g]=
\dfrac{1}{3} \sum_{x,\mu} \re \tr \left( g_{x}^{\dagger} U_{x\mu} g_{x+\mu} \right)
\label{eq:functional}
\eeq
by successive local gauge transformations $g_{x}$ acting on the link variables as 
follows
\beq
U_{x\mu} \stackrel{g}{\mapsto} U_{x\mu}^{g}
= g_x^{\dagger} U_{x\mu} g_{x+\mu} \,,
\qquad g_x \in SU(3) \,.
\label{gaugetraf}
\eeq
In order to achieve this, we subsequently apply two methods, first simulated
annealing ($SA$) and then over-relaxation ($OR$). $OR$ is applied to finally
satisfy the gauge condition (\Eq{eq:gaugecondition}) with a local accuracy of
\beq
\max_{x}\re\tr[\nabla_{\mu}A_{x\mu}\nabla_{\nu}A_{x\nu}^{\dagger}]<
10^{-13} \,,
\eeq
while $SA$ to reduce the Gribov ambiguity of lattice Landau gauge, by
favoring gauge-fixed (Gribov) copies with large values for
$F_{U}[g]$,
see~\cite{Parrinello:1990pm,Zwanziger:1990tn,Bakeev:2003rr,Sternbeck:2005tk,
Bogolubsky:2005wf,Bogolubsky:2007bw,Bornyakov:2009ug,Bogolubsky:2009dc}. 
For this the $SA$ algorithm generates gauge transformations $\{g_{x}\}$ 
randomly by a Monte Carlo chain with a statistical weight 
$\sim \exp(F_{U}[g]/T_{sa})$. The ``simulated annealing temperature'' 
$T_{sa}$ is a technical parameter which is monotonously lowered. Our annealing
schedule is specified by a hot start at $T_{sa}=0.45$, after which $T_{sa}$ is
continuously lowered in equal steps until $T_{sa}=0.01$ is reached.
We apply 3500 $SA$ steps between these two temperatures, and, for better
performance, also added a few microcanonical steps to each (heatbath) step.
Since we apply large number of $SA$ steps to maximize $F_{U}[g]$, we restrict
ourselves to one Gribov copy per configuraton.

Our first quantity of interest is the gluon propagator defined in momentum 
space as the ensemble average
\beq
 D^{ab}_{\mu\nu}(q) 
 =\left\langle \widetilde{A}^a_{\mu}(k)\widetilde{A}^b_{\nu}(-k) \right\rangle,
\label{eq:gluonzero}
\eeq
where $\langle\cdots\rangle$ represents the average over configurations.  
$\widetilde{A}^a_{\mu}(k)$ denotes the Fourier transform of the gauge potential 
(\ref{eq:potential}) and $k_\mu\in \left(-N_{\mu}/2, N_{\mu}/2\right]$ is the
lattice momentum ($\mu=1,\ldots,4$), which relates to the physical
momentum $q_\mu$ as 
\beq
q_{\mu}(k_{\mu}) = \frac{2}{a} \sin\left(\frac{\pi
 k_{\mu}}{N_{\mu}}\right)\,.
\eeq
Henceforth we will use the notation $(N_{\sigma}; N_{\tau})\equiv(N_i;
N_4)$ where $i=1,2,3$. 

For non-zero temperature Euclidean invariance is broken, and it is useful to split 
$D^{ab}_{\mu\nu}(q)$ into two components, the transversal $D_T$
(``chromomagnetic'') and the longitudinal $D_L$ (``chromoelectric'') propagator,
respectively,
\beq
D^{ab}_{\mu\nu}(q)=\delta^{ab} 
            \left (P^{T}_{\mu\nu} D_{T}(q_{4}^{2},\vec{q}^{\,2})+
                   P^{L}_{\mu\nu} D_{L}(q_{4}^{2},\vec{q}^{\,2}) \right ) \; .
\eeq
The fourth momentum component $q_4$ conjugate to the Euclidean time 
(Matsubara frequency) will be restricted to zero later on.
For Landau gauge $P^{T,L}_{\mu\nu}$ represent projectors transversal 
and longitudinal relative to the time-direction $(\mu=4)$:
\bea
P^{T}_{\mu\nu}&=&(1-\delta_{\mu4})(1-\delta_{\nu4}) 
       \left(\delta_{\mu\nu}- \frac{q_{\mu}q_{\nu}}{\vec{q}^{\;2}}\right), \\
P^{L}_{\mu\nu}&=&\left(\delta_{\mu\nu}-\frac{q_{\mu}q_{\nu}}{\vec{q}^2}\right)
                -P^{T}_{\mu\nu}\;.
\eea
For the propagators $D_{T,L}$ [or their respective dimensionless 
dressing functions  $Z_{T,L}(q)=q^2 D_{T,L}(q)$] we find
\beq
 D_T(q)=\frac{1}{2 N_g} 
        \left\langle \sum_{i=1}^3  
         \widetilde{A}^a_i(k) \widetilde{A}^a_i(-k)
        -\frac{q_4^2}{\vec{q}^{\;2}} 
        \widetilde{A}^a_4(k)\widetilde{A}^a_4(-k) \right\rangle
\eeq
and
\beq
 D_L(q)= \frac{1}{N_g}\left(1 + \frac{q_4^2}{\vec{q}^{\;2}}\right) 
        \left\langle \widetilde{A}^a_4(k) \widetilde{A}^a_4(-k) \right\rangle
\; ,
\eeq
where $N_g=N_c^2-1$ and $N_c=3$.
The zero-momentum propagator values are then defined as
\bea
\label{eq:zeromomprop}
D_T(0) &=& \frac{1}{3 N_g}
     \sum_{i=1}^3 \left\langle \widetilde{A}^a_i(0) \widetilde{A}^a_i(0)
\right\rangle, 
\\
D_L(0) &=& \frac{1}{N_g}
       \left\langle \widetilde{A}^a_4(0) \widetilde{A}^a_4(0) \right\rangle \;. 
\eea
Note that we have neglected a possible $O(a)$ improvement related to the use of
the improved gauge action \Eq{sym_action}.

The Landau gauge ghost propagator is given by
\bea
\label{eq:ghost} \nonumber
G^{ab}(q)&=&a^{2}\sum_{x,y}\langle e^{-2\pi i(k/N)\cdot(x-y)} [M^{-1}]^{ab}_{xy}\rangle \\
         &=&\delta^{ab}~G(q)= \delta^{ab}~J(q)/q^2\;,
\eea
where the four-vector $(k/N) \equiv (k_{\mu}/N_{\mu})$. $J(q)$ denotes the ghost
dressing function. The matrix $M$ is the lattice Faddeev-Popov operator
\beq
M^{ab}_{xy}=\sum_{\mu}[A^{ab}_{x,y}\delta_{x,y}-
                       B^{ab}_{x,y}\delta_{x+\hat{\mu},y}-
                       C^{ab}_{x,\mu}\delta_{x-\hat{\mu},y}]
\eeq
with
\begin{align*}
 A^{ab}_{x,y}&=\quad \re \tr [\{T^{a},T^{b}\}(U_{x,\mu}+U_{x-\hat{\mu},\mu})], \\
 B^{ab}_{x,y}&=2\cdot \re \tr [T^{b} T^{a} U_{x,\mu}],\\
 C^{ab}_{x,y}&=2\cdot \re \tr [T^{a} T^{b} U_{x-\hat{\mu},\mu}] \; ,
\end{align*}
written in terms of $T^{a}$, $a= 1, \ldots, N_g$, i.e. the Hermitian generators
of 
the \textit{su(3)} Lie algebra normalized according to 
$\tr [T^{a} T^{b}]=\delta^{ab}/2$. For the inversion of $M$ we use the
pre-conditioned conjugate gradient algorithm of \cite{Sternbeck:2005tk} with
plane-wave sources $\vec{\psi}_{c}$ with color and position components 
$\psi^{a}_{c}(x)=\delta^{a}_{c}\exp(2\pi\,ik\cdot(x/N))$.

To reduce lattice artifacts, we apply cylinder and cone cuts to our
data \cite{Leinweber:1998uu}. Specifically we consider only diagonal 
and slightly off-diagonal momenta for the gluon propagator and diagonal
momenta for the ghost propagator. Moreover, only modes with 
zero Matsubara frequency ($k_4=0$) are used.

\section{Gluon and ghost propagator results}
\label{sec:results_propagators}

\subsection{Momentum dependence}

\begin{figure*}[tb]
\centering
\mbox{
 \includegraphics[angle=0,scale=0.65]{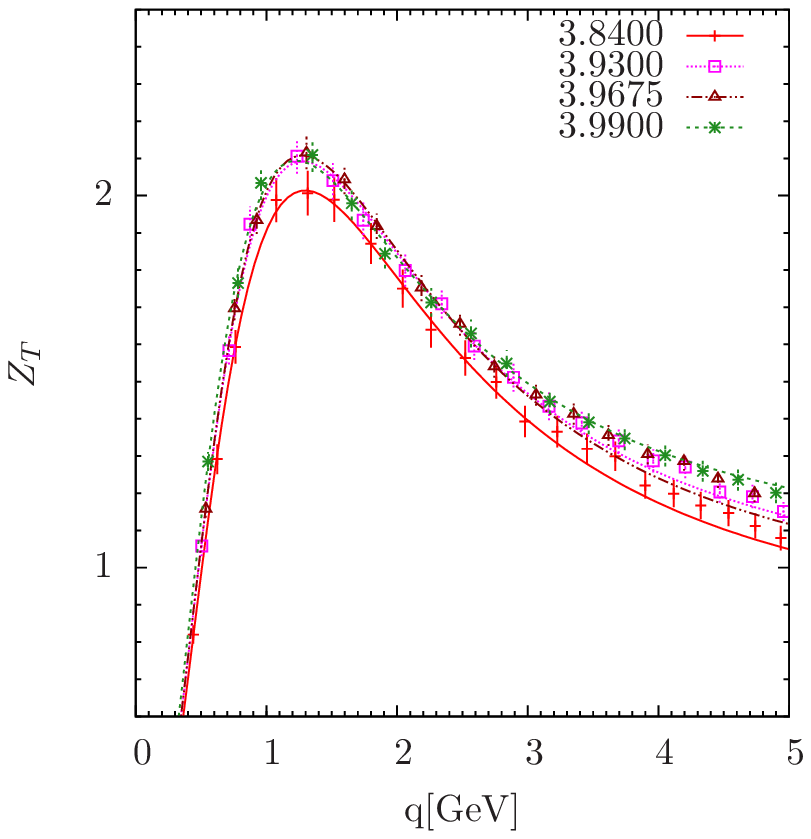} 
\hspace*{0.5cm}
 \includegraphics[angle=0,scale=0.65]{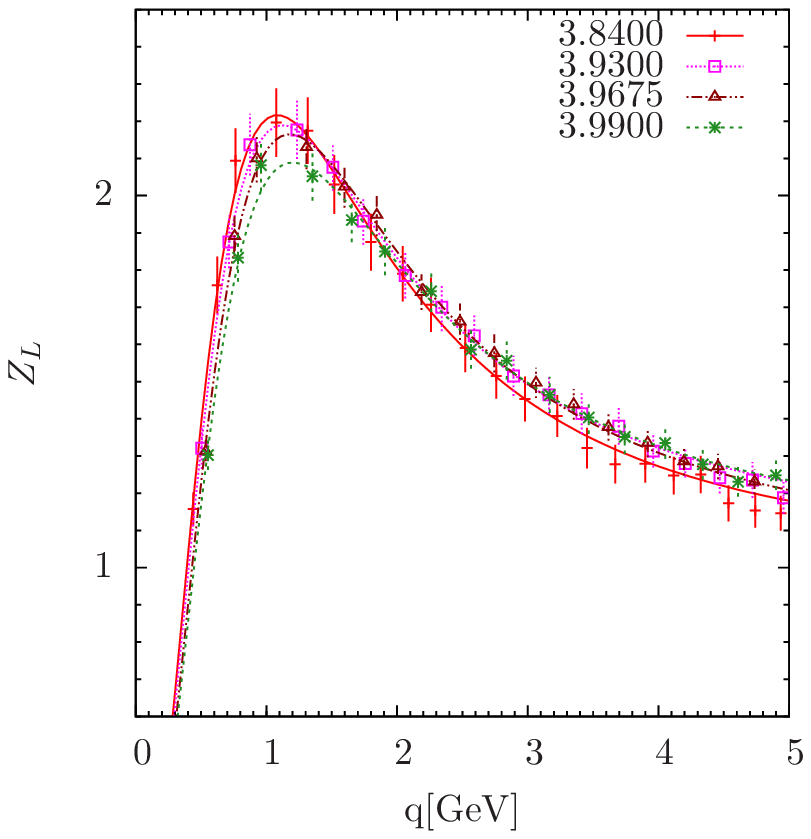}
\hspace*{0.5cm}
 \includegraphics[angle=0,scale=0.65]{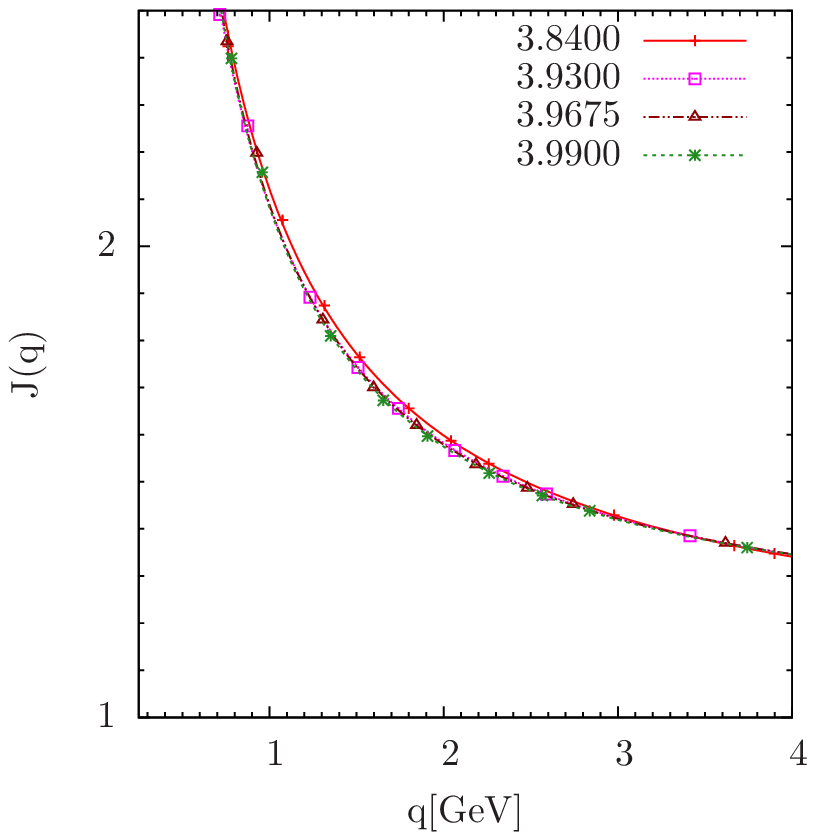}
}
\centering
 \mbox{
 \includegraphics[angle=0,scale=0.65]{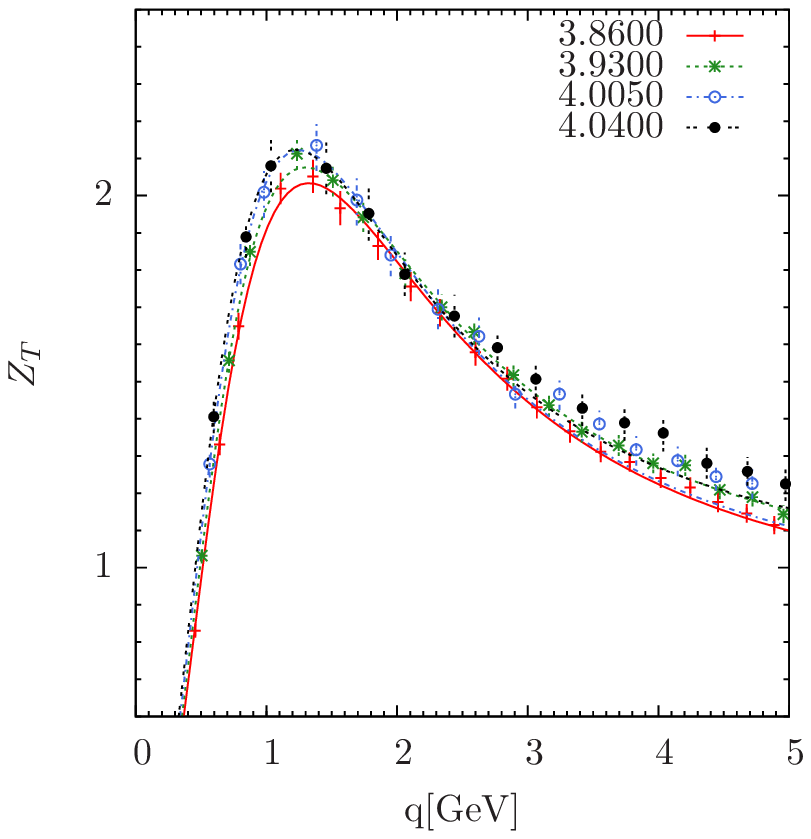} 
\hspace*{0.5cm}
 \includegraphics[angle=0,scale=0.65]{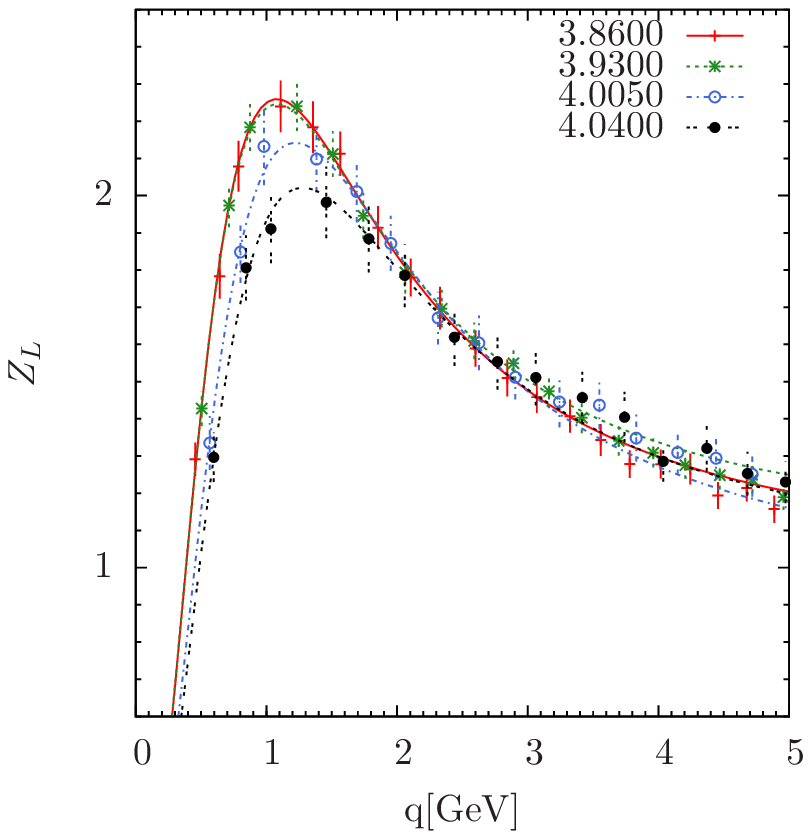} 
\hspace*{0.5cm}
 \includegraphics[angle=0,scale=0.65]{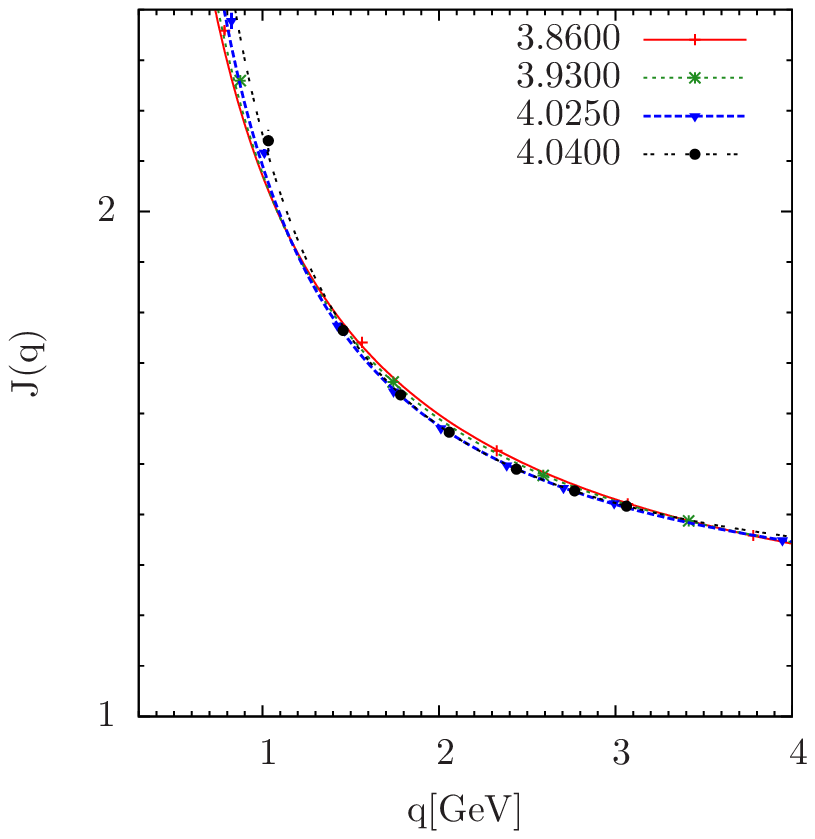}
 }
\centering
 \mbox{
 \includegraphics[angle=0,scale=0.65]{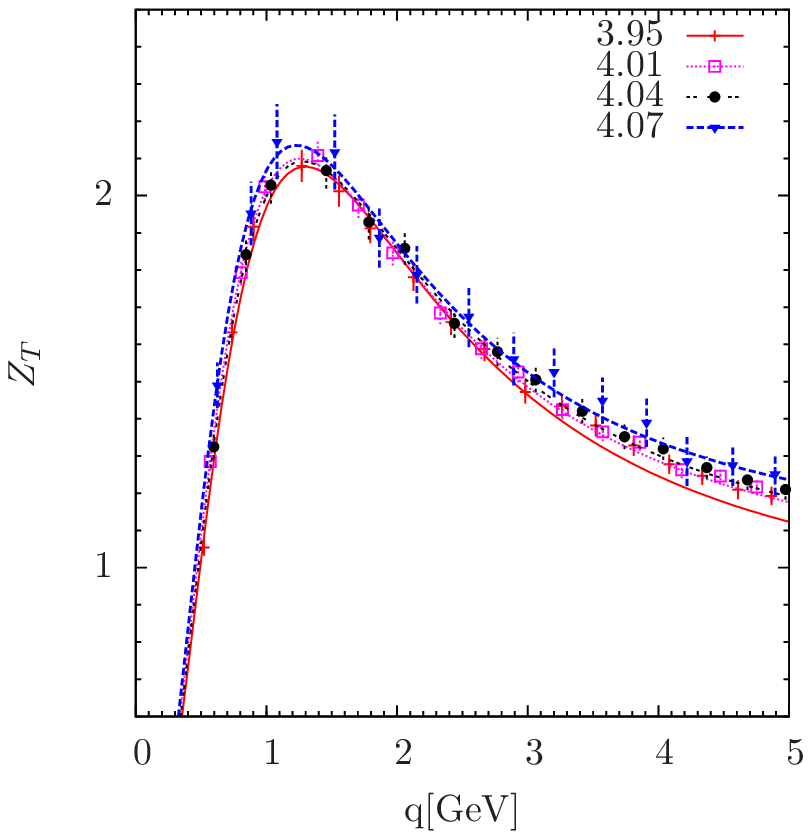} 
\hspace*{0.5cm}
 \includegraphics[angle=0,scale=0.65]{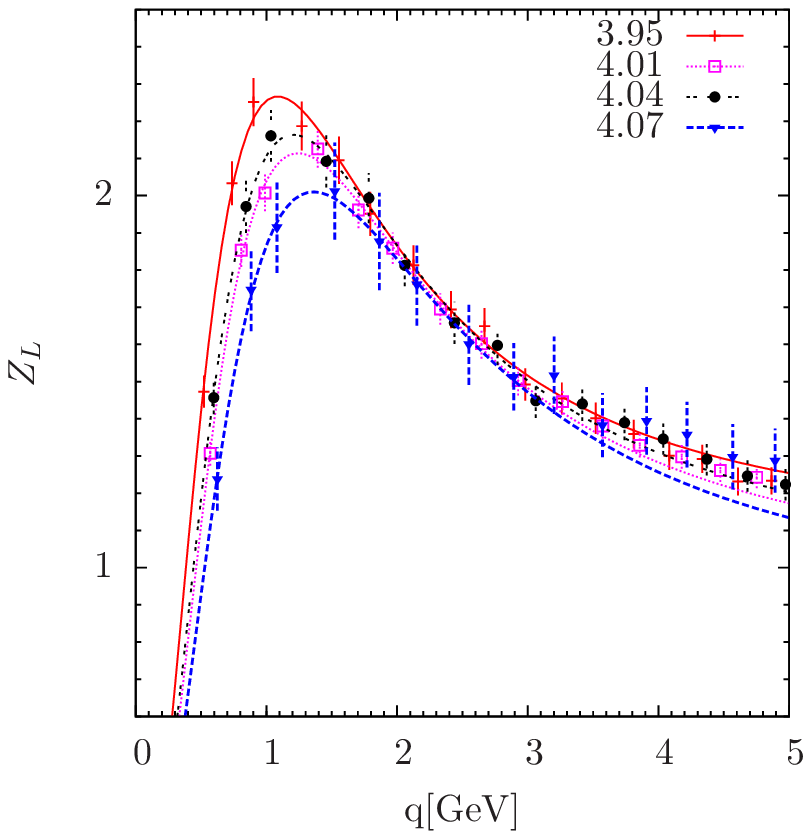} 
\hspace*{0.5cm}
 \includegraphics[angle=0,scale=0.65]{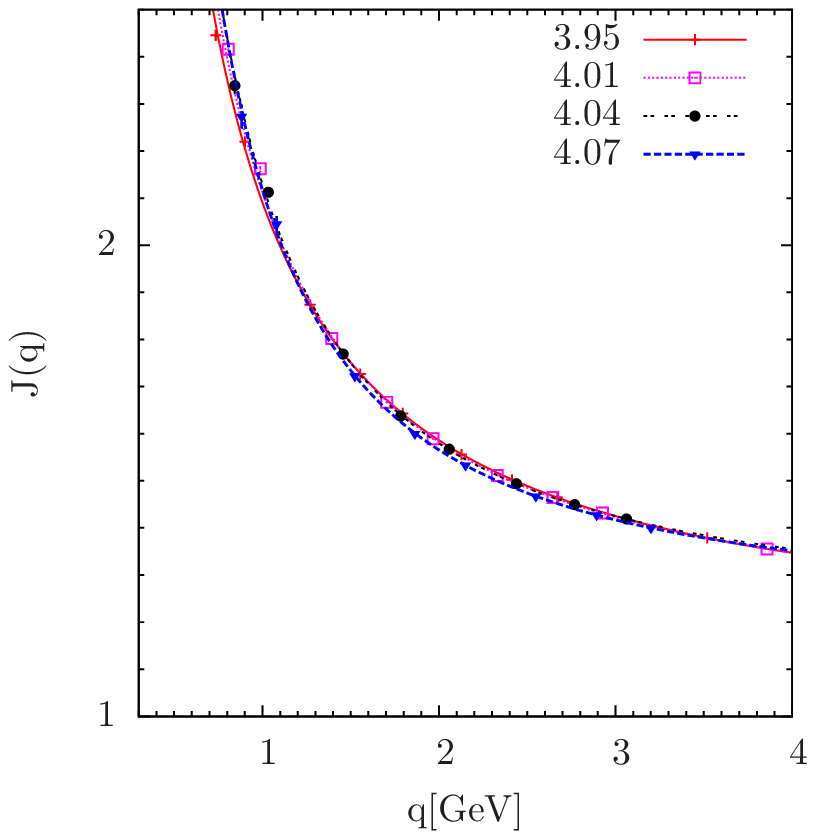}
 }
\caption{The unrenormalized dressing functions for the transverse gluon 
$Z_{T}$ (left panel), for the longitudinal gluon $Z_{L}$ (middle panel) 
and for the ghost dressing function $J$ (right panel) are shown as functions 
of the momentum $q~\mathrm{[GeV]}$ for different (inverse) coupling values 
$\beta$ (i.e. different temperatures) given in the legend. The corresponding 
pion mass values (from top to bottom panels) are $m_{\pi} \simeq 316, ~398$, 
and $469~\mathrm{MeV}$.}
\label{fig:dressfct_vs_q}
\end{figure*}

Data for the unrenormalized transverse ($Z_T$) and longitudinal ($Z_L$) gluon
dressing functions and also for the ghost dressing function ($J$) is shown in
\Fig{fig:dressfct_vs_q}. We show it versus the physical momentum $q \equiv
|\vec{q}|$ for selected temperatures and for three pion masses (panels
from top to bottom are for $m_\pi \simeq 316$, 398 and 469\,MeV, respectively).
The corresponding renormalized functions, in momentum subtraction (MOM) schemes,
can be obtained from
\bea
Z_{T,L}^{ren}(q,\mu) &\equiv& \tilde{Z}_{T,L}(\mu) Z_{T,L}(q),  \nonumber \\
J^{ren}(q,\mu) &\equiv& \tilde{Z}_{J}(\mu) J(q) \label{eq:renormZ}
\eea
with the $\tilde{Z}$-factors being defined such that
$Z_{T,L}^{ren}(\mu,\mu) =  J^{ren}(\mu,\mu) = 1$. For a renormalization
scale of $\mu=2.5~\mathrm{GeV}$ the $\tilde{Z}$-factors are quoted in
\Tab{tab:numbers}. 

\Fig{fig:dressfct_vs_q} also shows curves connecting data points of
same temperature. These were obtained from fits to the data for
momenta $0.4\,\mathrm{GeV} \le q \le 3.0\,\mathrm{GeV}$. These curves may serve
as input to studies of the corresponding DS or FRG equations. 

More specifically, for the gluon dressing function we employed (analogously to
our quenched study \cite{Aouane:2011fv}) the Gribov-Stingl formula 
\cite{Gribov:1977wm,Stingl:1994nk}\beq
\label{eq:fitgluon}
Z_\mathrm{fit}(q)= q^2 \frac{c~(1+d\,q^{2n})}{(q^2+r^2)^2+b^2}\;,
\eeq
which has been also used in \cite{Cucchieri:2003di,Cucchieri:2011di} and
appears in the context of the so-called ``Refined
Gribov-Zwanziger'' framework \cite{Dudal:2008sp,Dudal:2011gd}. But we found 
it sufficient to set $b^2=0$ and $n=1$. This fits well the data for 
$q\in[0.4,3.0]\,\mathrm{GeV}$ and gives excellent $\chi^2_{dof}$
values. The latter together with the results for the fit parameters are
listed in \Tab{tab:fits}. Note again that \Fig{fig:dressfct_vs_q} shows
data and the corresponding fits only for a selected range of temperatures, but
\Tab{tab:fits} gives the fit parameters for all available temperatures. We
cannot exclude that the $b^2$-term is needed for smaller momenta. If true, it
would indicate the occurrence of a pair of complex-conjugate poles.

For momenta above 3 GeV the fit fails to describe the data. In this range
logarithmic corrections are expected to become important. 

For the ghost dressing function we propose to use a fit formula like
\beq
J_\mathrm{fit}(q)=\left ( \frac{f^2}{q^2}\right)^k + h
\label{eq:ghostfit}
\eeq
In a first attempt we also tried $hq^2/(q^2+m_{gh}^2)$ for the last term with
$m_{gh}$ as a free parameter, but this was always found being consistent with
$m_{gh}=0$. We therefore omit such infrared mass parameter and only keep
a constant term $h$ in the ultraviolet limit. 

Fit results for the fitting range $[0.4\,\mathrm{GeV}, 4.0\,\mathrm{GeV}]$
are presented also in \Tab{tab:fits}. One notes that our $\chi^2_{dof}$ values
are far from being optimal, in particular for the lower temperatures.
Deviations typically occur at the lowest momenta. But this could not be cured,
e.g., by a mass term $m_{gh}^2$ alone. However note, the maximal deviations
of fit and data points do not exceed $5\%$.

\begin{table*}
{\scriptsize
\mbox{
\setlength{\tabcolsep}{1.0pt}
\begin{tabular}{|c|c|c|c|c|}  
\hline
 & \multicolumn{4}{|c|}{$Z_T$ fits} \\
\hline
 $\beta$ & $~c/a^2~$ & $~d/a^2~$ & $~a~r~$ & $~\chi^2_{dof}~$ \\
\hline
3.8400 & 1.868(142) & 0.420(78) & 0.510(12)  & 0.13 \\
3.8800 & 1.729(60) & 0.463(40) & 0.486(5)  & 0.66 \\
3.9300 & 1.371(104) & 0.647(111) & 0.431(10)  & 0.36 \\
3.9525 & 1.218(76) & 0.757(96) & 0.411(8)  & 0.22 \\
3.9600 & 1.208(88) & 0.744(122) & 0.405(10)  & 0.05 \\
3.9675 & 1.256(83) & 0.682(107) & 0.410(9)  & 0.10 \\
3.9750 & 1.099(93) & 0.852(156) & 0.387(11)  & 0.07 \\
3.9900 & 0.982(70) & 1.026(143) & 0.369(9) & 0.38 \\
\hline
\hline
3.8600 & 1.811(109) & 0.459(66) & 0.501(9) &  0.32\\
3.8800 & 1.637(88) & 0.523(63) & 0.477(8)  & 0.66 \\
3.9300 & 1.389(84) & 0.651(86) & 0.437(8)  & 0.37 \\
3.9700 & 1.151(71) & 0.819(110) & 0.397(8)  & 0.48 \\
3.9900 & 1.035(171) & 0.965(310) & 0.383(21) & 0.25 \\
4.0050 & 1.092(104) & 0.784(162) & 0.381(12)  & 0.25 \\
4.0175 & 1.035(87) & 0.891(164) & 0.373(10) & 0.44 \\
4.0250 & 1.020(78) & 0.904(144) & 0.368(9) & 0.21 \\
4.0400 & 0.837(111) & 1.237(329) & 0.340(15) & 0.12 \\
\hline
\hline
3.9500 & 1.338(80) & 0.645(89) & 0.427(8)  & 0.63 \\
3.9700 & 1.290(104) & 0.648(127) & 0.416(11)  & 0.14 \\
3.9900 & 1.126(91) & 0.797(139) & 0.389(10) & 0.28 \\
4.0100 & 1.009(61) & 0.937(115) & 0.371(7)  & 0.32 \\
4.0200 & 1.027(67) & 0.946(124) & 0.375(8) & 0.07 \\
4.0300 & 1.096(79) & 0.775(135) & 0.380(9) & 0.48 \\
4.0400 & 0.941(84) & 1.018(191) & 0.359(11) & 0.22 \\
4.0500 & 1.053(69) & 0.781(126) & 0.367(8) & 0.67 \\
4.0700 & 0.792(130) & 1.291(405) & 0.328(34) & 0.15 \\
\hline
\end{tabular}
}
\hspace*{-0.3cm}
\mbox{
 \setlength{\tabcolsep}{3.0pt}
\begin{tabular}{|c|c|c|c|}
\hline
   \multicolumn{4}{|c|}{$Z_{L}$ fits} \\
\hline
$~c/a^2~$ & $~d/a^2~$ & $~a~r~$ & $~\chi^2_{dof}~$  \\
\hline
1.334(132) & 0.744(138) & 0.415(14)  & 0.19 \\
1.183(59) & 0.872(79) & 0.390(7)  & 0.16 \\
1.032(101) & 1.013(188) & 0.370(13)  & 0.12 \\
1.049(93) & 0.941(159) & 0.370(11)  & 0.16 \\
0.932(103) & 1.148(233) & 0.355(14)  & 0.13 \\
1.023(108) & 0.979(206) & 0.369(13)  & 0.27 \\
0.938(119) & 1.165(277) & 0.359(15)  & 0.09 \\
0.914(98) & 1.143(224) & 0.358(13) & 0.38 \\
\hline
\hline
1.271(116) & 0.799(143) & 0.401(13) & 0.19 \\
1.218(83) & 0.808(107) & 0.391(9)  & 0.11 \\
0.982(91) & 1.092(177) & 0.356(12) & 0.11 \\
1.034(86) & 0.928(160) & 0.366(10)  & 0.67 \\
1.171(265) & 0.699(430) & 0.383(42) & 0.21 \\
1.006(145) & 0.924(281) & 0.366(19)  & 0.15 \\
0.922(133) & 1.087(297) & 0.356(18) & 0.12 \\
0.845(94) & 1.278(270) & 0.346(13)  & 0.11 \\
0.824(173) & 1.252(502) & 0.348(24) & 0.08 \\
\hline
\hline
0.952(78) & 1.127(169) & 0.348(10)  & 0.35 \\
0.993(100) & 1.003(199) & 0.351(12)  & 0.32 \\
0.883(96) & 1.203(236) & 0.343(13)  & 0.19 \\
0.998(82) & 0.945(157) & 0.368(10)  & 0.23 \\
1.010(80) & 0.918(156) & 0.356(9)  & 1.08 \\
0.900(87) & 1.108(223) & 0.349(11) & 0.67 \\
0.870(103) & 1.131(275) & 0.340(14) & 0.25 \\
0.826(89) & 1.251(260) & 0.344(13) & 0.25 \\
0.971(250) & 0.894(559) & 0.371(32) & 0.01 \\
\hline
\end{tabular}
}
\hspace*{-0.3cm}
\mbox{
 \setlength{\tabcolsep}{3.0pt}
\begin{tabular}{|c|c|c|c|}
\hline
\multicolumn{4}{|c|}{$Z_J$ fits} \\
\hline
 $a^2 f^2$ & $h/a^2$ & $k$ & $\chi^2_{dof}$  \\
\hline
 0.4580(17) & 1.0916(61) & 0.5111(78) & 0.69 \\ 
 0.41822(7) & 1.0904(19) & 0.4950(23) & 8.59 \\ 
 0.37046(9) & 1.1355(39) & 0.5438(55) & 2.29 \\
 0.35672(9) & 1.1387(36) & 0.5462(52) & 2.40 \\
            &            &            &      \\
 0.34636(7) & 1.1501(33) & 0.5642(54) & 5.20 \\
            &            &            &      \\
 0.33093(8) & 1.1571(30) & 0.5736(56) & 7.01 \\
\hline\hline
 0.4464(21) & 1.0419(40) & 0.4444(35)& 31.1 \\ 
            &            &            &      \\
 0.3802(19) & 1.0962(53) & 0.4945(58)& 21.4 \\ 
            &            &            &      \\
 0.3380(07) & 1.1466(25) & 0.5612(39)& 24.0\\
 0.441(84)  & 0.90(14)   & 0.41(06)  & 0.21\\
 0.3189(09) & 1.1623(31) & 0.5834(60)& 6.4 \\
 0.3155(08) & 1.1630(25) & 0.5853(52)& 17.4 \\
 0.3135(21) & 1.2200(90) & 0.696(21) & 0.93 \\
\hline\hline
 0.3621(09) & 1.1325(38) & 0.5397(53) & 2.3 \\
 0.3519(07) & 1.1426(31) & 0.5554(48) & 4.2 \\
 0.3405(12) & 1.1555(45) & 0.5787(81) & 2.3 \\
 0.3295(07) & 1.1607(24) & 0.5873(48) & 13.1 \\
 0.3270(08) & 1.1557(27) & 0.5778(51) & 7.0 \\
 0.3130(08) & 1.1538(26) & 0.5660(48) & 7.9 \\
 0.3100(11) & 1.2015(55) & 0.648(12)  & 2.3\\
 0.3082(11) & 1.2109(51) & 0.677(12)  & 0.94\\
 0.2913(20) & 1.2156(79) & 0.682(19)  & 0.39\\
\hline
\end{tabular}
}
\caption{Results from fits with the Gribov-Stingl formula \Eq{eq:fitgluon}
for the unrenormalized $Z_T$ (left subtable) and $Z_L$ (center subtable) gluon
dressing functions. The fit range is $[0.4 : 3.0]~\mathrm{GeV}$.
The values in parentheses indicate the fit errors estimated with the bootstrap
method. The parameters $b$ and $n$ were fixed to $b=0$ and $n=1$, respectively. \\
Fit results for the unrenormalized ghost dressing function $Z_J$
with the fitting function according to \Eq{eq:ghostfit} are presented in the
right subtable. The momentum fitting ranges are here $[0.4 : 4.0]$ GeV.
The pion mass values are $m_{\pi}=316 (16)~\mathrm{MeV}$ (upper), 
$m_{\pi}=398 (20)~\mathrm{MeV}$ (middle) and $m_{\pi}=469 (24)~\mathrm{MeV}$ 
(bottom subtables), respectively. 
}
\label{tab:fits}
}
\end{table*}

\subsection{Temperature dependence}

We now look at the temperature dependence of the dressing functions, where
our temperature values cover the chiral restoration and the deconfinement phase
transition, with the latter being signaled by a peak in the Polyakov loop
susceptibility. These two crossover phenomena typically occur at different
temperatures and will be denoted $T_\chi$ and $T_\mathrm{deconf}$, respectively,
in what follows (see \Tab{tab:tc} for their values). 

Looking once again at \Fig{fig:dressfct_vs_q}, we see that the momentum
dependences of $Z_L(q)$ and $Z_T(q)$ change differently with temperature,
irrespective of $m_\pi$. In fact, while the (unrenormalized) transverse dressing
function $Z_T$ seems to be relatively insensitive to the temperature, the curves
describing $Z_L(q)$ fan out for momenta below the renormalization scale
$\mu=2.5~\mathrm{GeV}$. A stronger temperature dependence we also
observed for $D_L(q)$ for pure $SU(3)$ gauge theory \cite{Aouane:2011fv}, though
there it was found to be much more pronounced due to the existence of a first
order phase transition 
\cite{Aouane:2011fv}.    

\begin{figure*}[tb]
 \centering
 \mbox{
 \includegraphics[angle=0,scale=0.8]{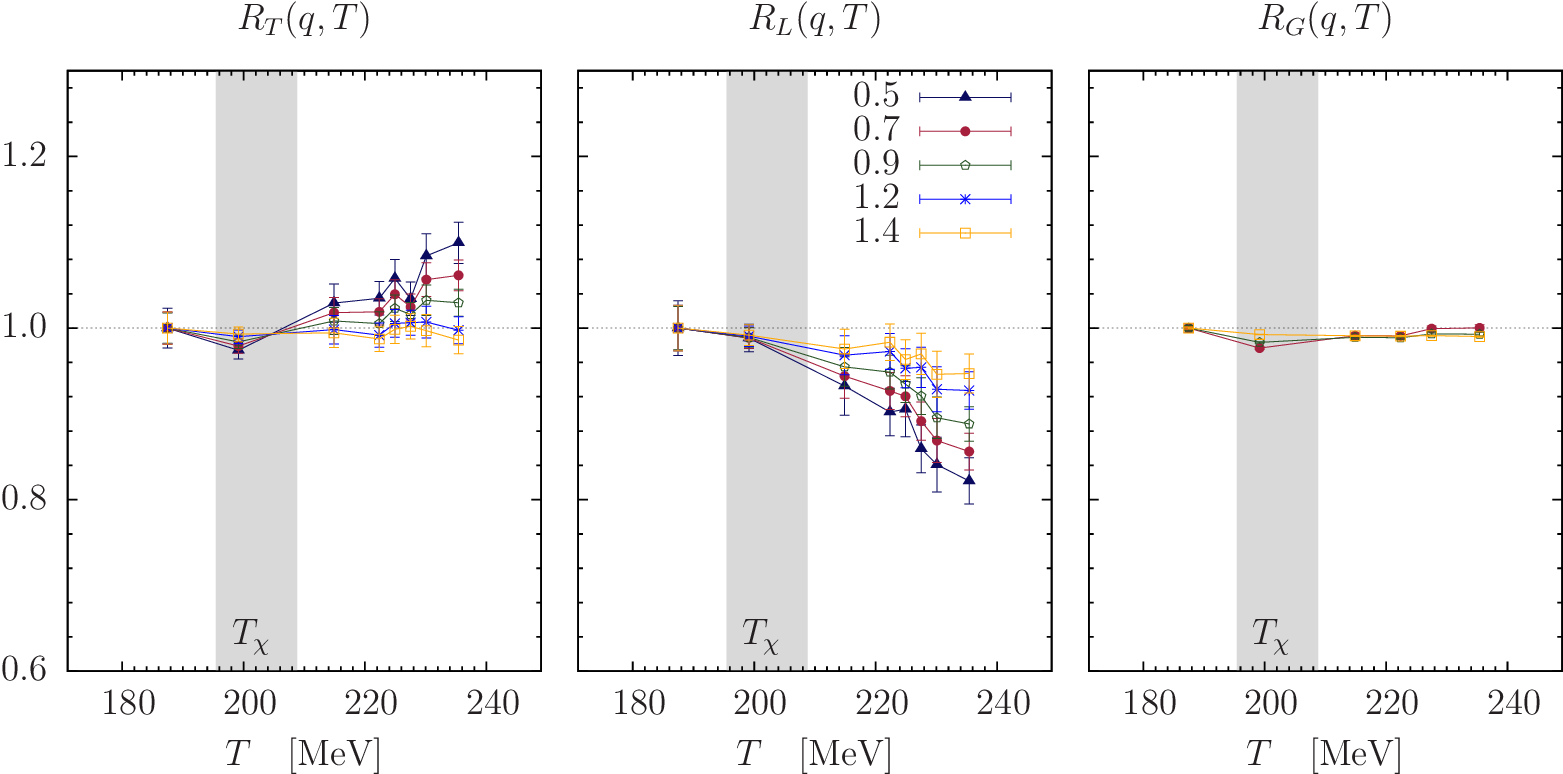} 
 }
 \par\vspace*{0.7cm}
 \centering
 \mbox{
 \includegraphics[angle=0,scale=0.8]{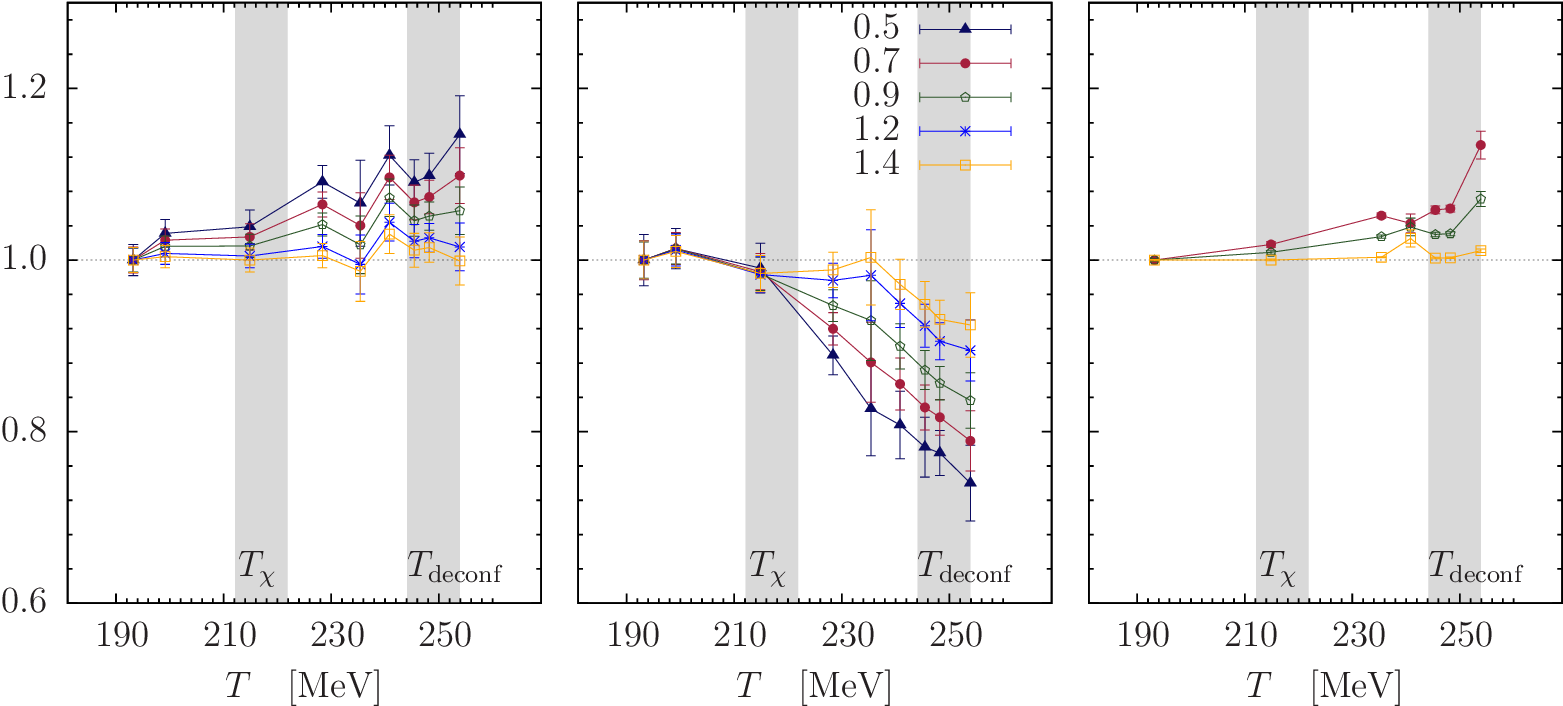} 
 }
 \par\vspace*{0.7cm}
\centering
 \mbox{
 \includegraphics[angle=0,scale=0.8]{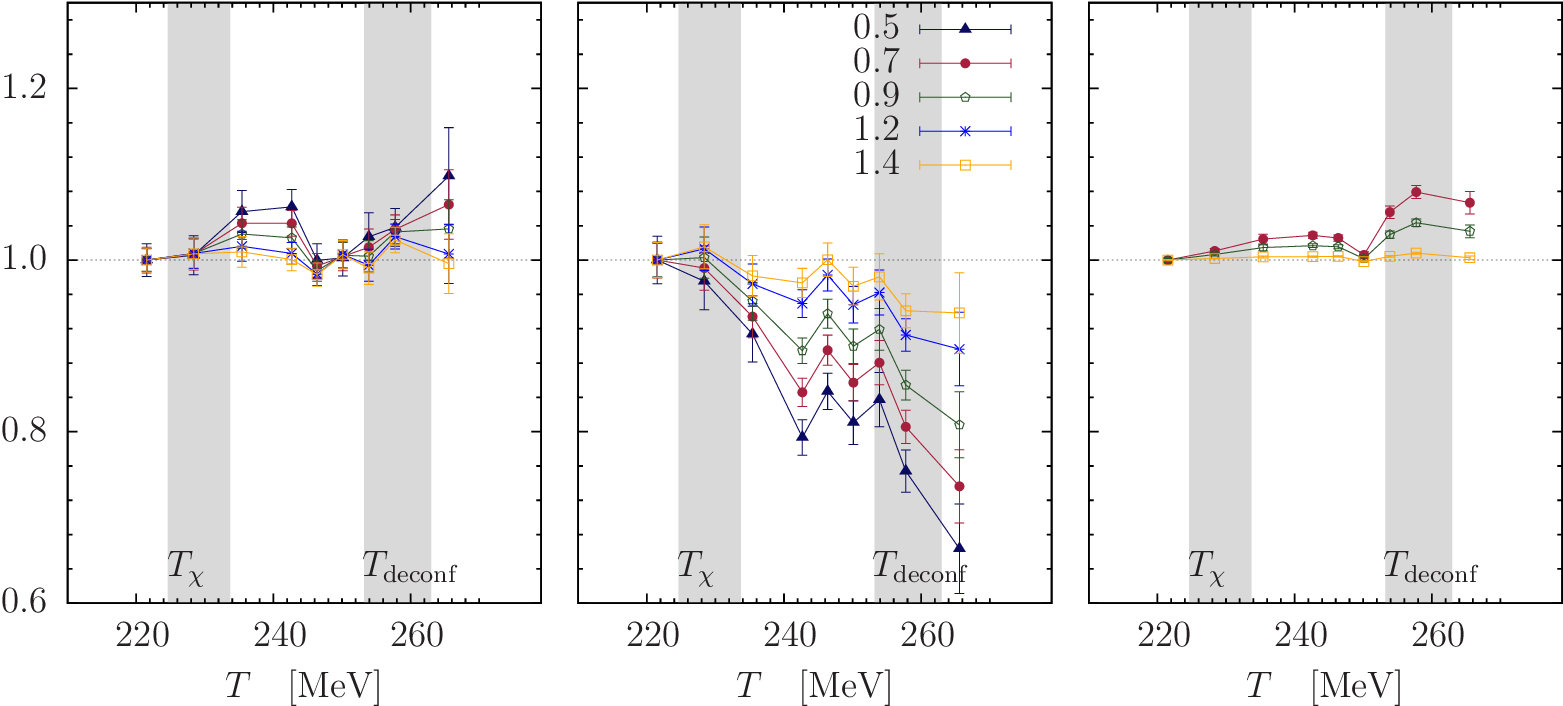} 
 }
\caption{Ratios $R_T, R_L$ and $R_G$ for the renormalized 
transverse $D_{T}^{ren}$ (left panel), longitudinal 
$D_{L}^{ren}$ (middle panel) and ghost $G^{ren}$ (right panel) propagators, 
respectively, as  functions of the temperature $T$ at a few non-zero
momentum values $q$ (indicated in units of $\mathrm{[GeV]}$. The corresponding 
pion masses (from top to bottom) are $m_{\pi} \simeq 316,~398$ and 
$469~\mathrm{MeV}$. The vertical bands indicate the chiral
and deconfinement pseudo-critical temperatures with their uncertainties 
(see \Tab{tab:tc}).}  
\label{fig:props_vs_T}
\end{figure*}

These observations are seen more clearly in \Fig{fig:props_vs_T}, 
where we show ratios of the renormalized dressing functions or 
propagators
\bea
R_{T,L}(q,T) &=& D_{T,L}^{ren}(q,T) / D_{T,L}^{ren}(q,T_\mathrm{min}), \\
R_G(q,T) &=& G^{ren}(q,T)/G^{ren}(q,T_\mathrm{min}) 
\eea
as functions of the temperature $T$ for 6 fixed (interpolated) momentum values 
$q \ne 0$, and for different pion masses (panels from top to bottom). For better
visibility, ratios are normalized with respect to the
respective left-most shown temperature in \Fig{fig:props_vs_T}.

Looking at \Fig{fig:props_vs_T}, we see $R_L(q,T)$ decrease more or less 
monotonously with temperature in the crossover region, and this
decrease is stronger the smaller the momentum. $R_T(q,T)$ instead signals a
slight increase within the same range, and the ghost propagator (at 
fixed low momenta) seems to rise a bit around $T \simeq T_\mathrm{deconf}$.

\begin{figure*}[tb]
 \centering
 \mbox{
 \includegraphics[angle=0,scale=0.8]{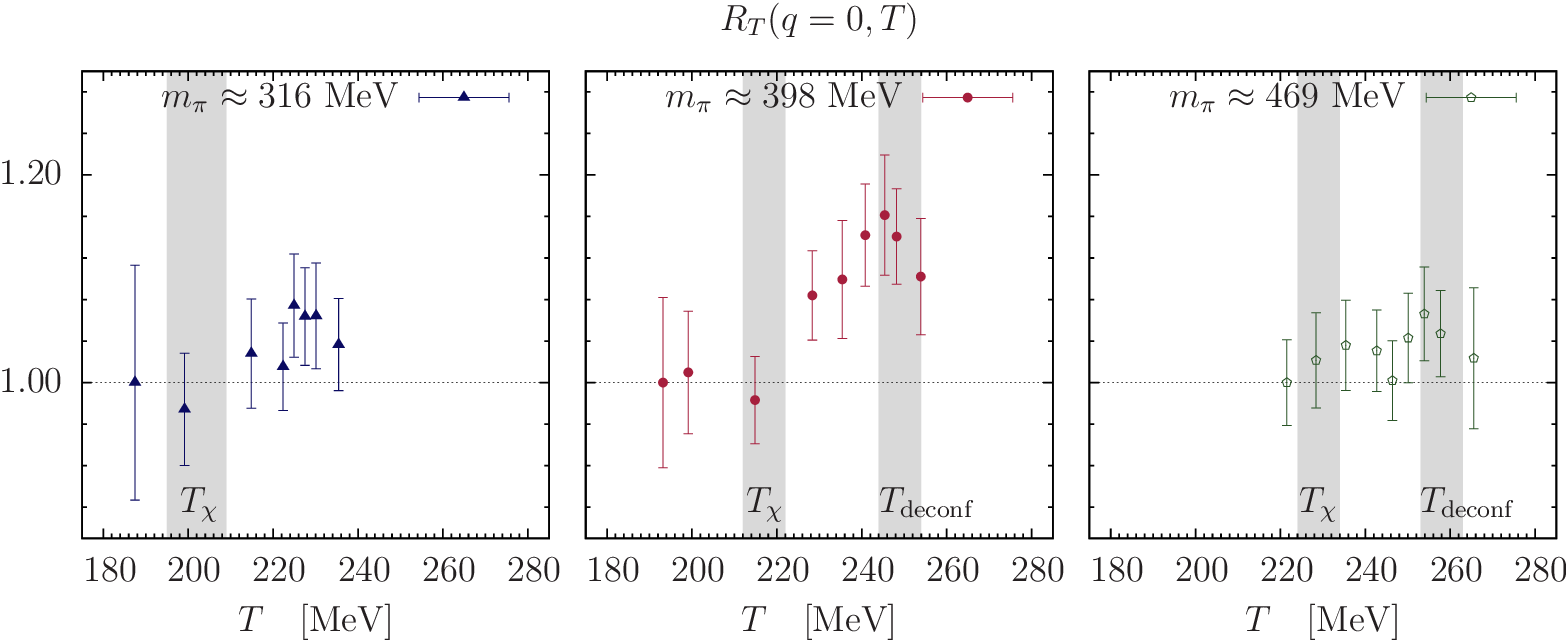} 
 }
 \par\vspace*{0.7cm}
 \centering
 \mbox{
 \includegraphics[angle=0,scale=0.8]{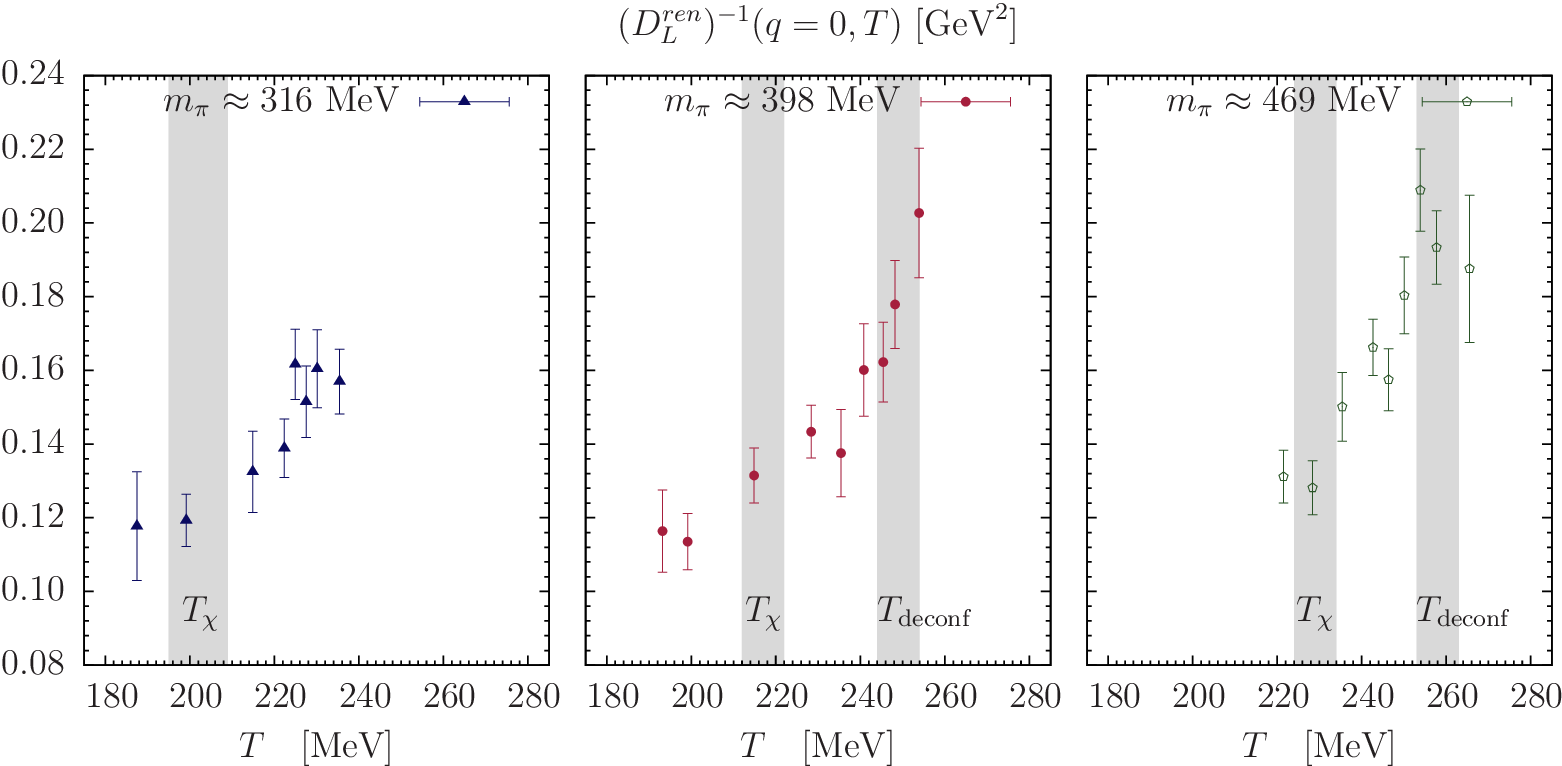} 
 }
\caption{The upper row shows the ratio $R_T$ at zero momentum for the three
pion mass values indicated. The lower panels show the inverse renormalized 
longitudinal gluon propagator $(D_L^{ren})^{-1}$ at zero momentum.}
\label{fig:gluon_zero_momentum}
\end{figure*}

\Fig{fig:props_vs_T} does not show ratios at zero momentum, but we show it for
$R_T$ in \Fig{fig:gluon_zero_momentum} (upper row), again versus $T$ and from
left to right for different pion masses. For $m_\pi\approx 398\,\text{MeV}$, for
example, $R_T(0,T)$ clearly rises towards $T_\mathrm{deconf}$, whereas there are
only weak indications for such a behavior for $R_T(0,T)$ for the other two
data sets. Much more statistics is necessary to resolve that. 

The lower panels of \Fig{fig:gluon_zero_momentum}, show data for the inverse
renormalized longitudinal propagator $D^{ren}_L$ at zero momentum, again versus
temperature and from left to right for different pion masses. This quantity can
be identified with an infrared gluon screening mass, and we clearly see it to
rise with temperature in the crossover region. This again shows that this
infrared gluon screening mass may serve as an useful indicator for the
finite-temperature crossover of the quark-gluon system. However, we should keep
in mind that the zero-momentum results for the gluon propagators are influenced
also by strong finite-size and Gribov copy effects, which we could not
analyze here.
 
\section{Conclusion}
\label{sec:conclusion}

We have presented data for the Landau gauge gluon and ghost propagators 
for lattice QCD at finite temperature with $N_f=2$ twisted mass fermion flavors.
Our data is for a momentum range of 0.4\,GeV to 4.0\,GeV and was obtained on
gauge field configurations produced by the tmfT Collaboration. This has
allowed us to explore the propagator's momentum dependence over the whole
temperature range of the crossover region, and this separately for three
(charged) pion mass values between 300\,MeV and 500\,MeV. We find that the
propagators change smoothly passing through the crossover region and the most
significant change is seen for the longitudinal (i.e., electric) component of
the gluon propagator.

We also provide fitting functions for our data. These and the corresponding fit
parameters, given in \Tab{tab:fits}, may serve as interpolation functions of our
data when used as input to studies which employ continuum functional methods to
address problems of QCD at finite temperature.
Actually, for both the transversal and longitudinal gluon dressing function
these interpolation functions give quite a good description
of our data for all temperatures. For the ghost dressing function, this is
achieved only for selected temperatures (see \Tab{tab:fits} for details). 

We hope our results will help these (continuum-based) studies to get further
predictions of the behavior of hadronic matter close to the transition region
that would be too difficult if addressed on the lattice directly. 

\section*{Acknowledgments}
We thank the members of the tmfT Collaboration for giving us access to their
gauge field configurations produced for $N_f=2$ fermion degrees of freedom
with the Wilson-twisted mass approach. We express our gratitude to 
the HLRN supercomputer centers in Berlin and Hannover for generous supply 
with computing time. 
R.~A. expresses thanks to L. Zeidlewicz for his help in carrying out the 
data analysis and gratefully acknowledges financial support 
by the Yousef Jameel Foundation. A.~S.\ acknowledges support from 
the European Reintegration Grant (FP7-PEOPLE-2009-RG No.256594),
and F.~B. from the DFG-funded graduate school GK 1504.

\bibliographystyle{apsrev}


\end{document}